\RequirePackage[2020/08/10]{latexrelease}
\documentclass[aps,pra,reprint,superscriptaddress,longbibliography]{revtex4-2}

\usepackage{physics}
\usepackage{amsmath}
\usepackage{amssymb}
\usepackage{graphicx}
\usepackage{float}
\usepackage{mathrsfs}
\usepackage{mathtools}
\usepackage{extarrows}
\usepackage[margin=0.7in]{geometry}
\graphicspath{{PNG/}}
\usepackage{xcolor}

\begin{document}
	
	\title{Asymmetrical post quench transport in an embedded parity time symmetric Su-Schrieffer-Heeger system}
	
	\author{A.~ Ghosh}
		
	\affiliation{School of Physics, University of Melbourne, Melbourne, 3010, Australia}
	
	\affiliation{Indian Institute of Technology Kharagpur, Kharagpur 721302, West Bengal, India}

	\author{A.~M.~Martin}
	
	\affiliation{School of Physics, University of Melbourne, Melbourne, 3010, Australia}

	\date{\today}

	\begin{abstract}
		
		We study the effect of PT-symmetric non-Hermiticity on the transport of edge state probability density arising as a result of a quench. A hybrid system involving a PT-symmetric SSH region sandwiched between two plain SSH systems is designed to study the dynamics. Geometrical arguments and numerical calculations were made to ascertain the nature of edge states. We then compute the quench dynamics numerically and demonstrate that the post-quench probability density light cones exhibit contrasting shapes as a result of asymmetrical reflections from the non-Hermitian part of the system depending on the direction of propagation of the transporting wave and, hence, on the initial localization of the edge state. 
		
	\end{abstract}
	
	\maketitle

	\section{Introduction \label{sec:1}}
The pioneering theoretical works of C.M. Bender and colleagues \cite{bender1998real, bender2005introduction, bender2007making} demonstrated that non-Hermitian operators can characterize real eigenvalues if they obey PT symmetry. Interesting theoretical works on the possibility of the physical implementation of the PT-symmetric Hamiltonian in optical settings followed ~\cite{klaiman2008visualization, makris2008beam, el2007theory, mostafazadeh2009spectral, lin2011unidirectional}. The last decade has seen many exotic phenomena involving PT-symmetric Hamiltonian and exceptional points~\cite{makris2008beam, lin2011unidirectional, el2018non, guo2009observation, ruter2010observation}. It has been shown that many of the exotic effects of PT-symmetric systems are not only limited to quantum systems but also to classical ones \cite{huber2016topological, anandwade2023synthetic}.
The exotic effects of PT-symmetric systems have been demonstrated on various platforms, including photonic systems~\cite{regensburger2012parity, ruter2010observation, miri2019exceptional}, atomic ensembles~\cite{zhang2016observation}, topoelectrical systems~\cite{gupta2021non}, mechanical systems\cite{huber2016topological}, and acoustic systems~\cite{fleury2015invisible}.
Some examples of these exciting effects of exceptional points in non-Hermitian PT-symmetric systems constitute unidirectional invisibility~\cite{lin2011unidirectional, regensburger2012parity, fleury2015invisible}, Bloch oscillations~\cite{longhi2009bloch}, and loss enhanced transmission through waveguides~\cite{guo2009observation}.

The concept of non-Hermiticity has also been implemented in the field of topological physics, and some interesting phenomenon has been observed in topological models perturbed with non-Hermitian elements~\cite{lieu2018topological, yao2018edge, zhang2020correspondence}. Interesting studies have been undertaken on the topology of quasiparticles with finite time that arise from the non-Hermiticity of the self-energy arising from electron-phonon coupling~\cite{kozii2017non}. It has been established that the Su-Schrieffer-Heeger(SSH) model~\cite{su1979solitons, asboth2016short}, which, under non-Hermitian conditions and obeying PT symmetry, is a periodic structure with alternating imaginary on-site potentials~\cite{lieu2018topological}. Very recently, it has been shown that an SSH system, after a quench from one topological regime to another, characterizes a post-quench transport that forms a light cone~\cite{PhysRevE.108.034102}.

Non-Hermiticity's impact on post-quench information spread or light cones needs to be explored, as little has been studied on this topic.
In this work, we shed light on the effect of non-Hermiticity with PT symmetry on the quench dynamics. We consider a hybrid system to investigate the impact of non-Hermiticity on post-quench transport. The system consists of a non-Hermitian PT-symmetric SSH system sandwiched between two Hermitian SSH systems. 
The hybrid system has been designed by considering that the edge states in the SSH system are localized at either end of the chain, and the post-quench transport takes place from the place of initial localization to the other end~\cite{PhysRevE.108.034102}. It is found that the quench dynamics is determined by the direction of propagation of the transport wave, and hence on the initial state corresponding to the quench since the initial state determines the direction of propagation. Depending upon the initial edge state, there can be two directions for the post-quench transport and we have found that for two directions of post-quench transport; in one case, there is negligible scattering from the middle PT-symmetric part, and in another, there is significant scattering, giving rise to a very different dynamics.
\\This paper is organized as follows. The first part of Section 2 introduces the schematics of the model. The Hamiltonian is introduced, and by geometric considerations, it is apprehended that there might be zero energy edge states at the edges of the hybrid system. In part B of this section, we numerically demonstrate the topological nature of the edge states by considering the stability of the spatial structure of the edge states by varying the hopping parameter. The energy band structure is also shown to indicate the topological nature corresponding to PT symmetry. We then move forward to the next section, where we discuss the results. In part A of this section, we perform a numerical analysis of the quench dynamics. Here, we study and compare the light cone diagrams for the two quenches with different initial edge states. In part B, we model the system using real and imaginary potential blocks and then proceed to calculate the scattering matrix considering a simple plane wave with fixed energy. Based on the analysis of the results, it is demonstrated that for different transport directions, there is a contrasting asymmetry in their properties. Conclusions and summaries were discussed in the final section.

	\section{Introduction to the system: Embeded non- Hermitian PT symmetric SSH \label{sec:2}}
	\subsection{Schematics of the Model }
\begin{figure}
	\centering
	\includegraphics[width=0.7\linewidth]{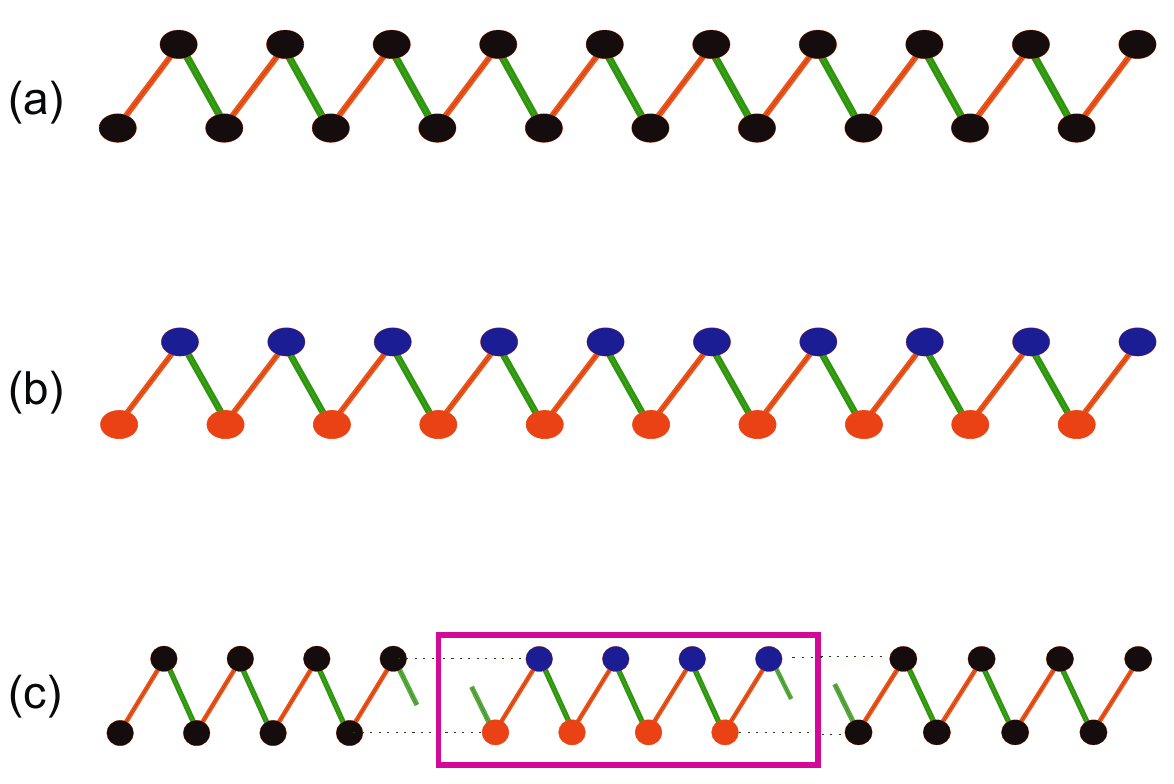}
	\caption{Schematics of (a) a plain SSH system, with staggered hopping amplitudes $ v $ and $ w $. (b) shows a non-Hermitian PT-symmetric SSH system with alternating complex on-site potentials, $ u $ and $ u^{*} $. The hybrid system comprising of a non-Hermitian PT symmetric SSH system sandwiched between two plain SSH systems is shown in (c).}
	\label{fig:figure1}
\end{figure}
	A SSH system can be made non-Hermitian by a number of ways \cite{lieu2018topological}. For example, the introduction of complex hopping amplitudes, keeping the onsite potentials at zero, introduces non-Hermiticity by preserving the chiral symmetry. On the other hand, a staggered complex onsite potential with real hopping amplitudes introduces non-Hermiticity whilst maintaining PT symmetry and breaking the chiral symmetry (refer to Figure 1(b)). In the non-hermitian PT-symmetric SSH model, two regions of the unbroken PT-symmetric phase are separated by an extended region of the broken PT-symmetric phase, making the topological transition very different from that of chiral symmetric ones \cite{lieu2018topological}.
We consider a hybrid system that comprises a PT-symmetric non-Hermitian SSH system sandwiched between two simple SSH systems as shown in Figure 1(c). The staggered potential on the middle PT symmetric part is complex, i.e., has the form $ u(u^{*}) = u_{Re}-(+)  iu_{Im} $. The Hamiltonian of our system is of the form:
\begin{equation}\label{eqn:1}
	\hat{H} = H_{SSH} + H_{PT},
\end{equation}
i.e.,
\begin{align}\label{eqn:2}
     \hat{H}&= v \Sigma_{n=1}^{N} \ket{A,n}\bra{B,n} + w \Sigma_{n=1}^{N-1} \ket{A,n}\bra{B,n+1} \\
	  &+ \Sigma_{n=n1}^{n=n2} u\ket{A,n}\bra{A,n} + u^{*}\ket{B,n}\bra{B,n}\nonumber.
\end{align}
In the Equation 2, $v$ represents the intracellular hopping, and $w$ represents the intercellular hopping; the non-Hermitian PT-symmetric region extends from the unit cell $n1$ to $n2$. Even without numerical consideration, a geometrical inspection of the system under the extreme dimerized case of $v = 0$ suggests the presence of zero energy edge states, as shown in Figure 2.
\begin{figure}
	\centering
	\includegraphics[width=0.7\linewidth]{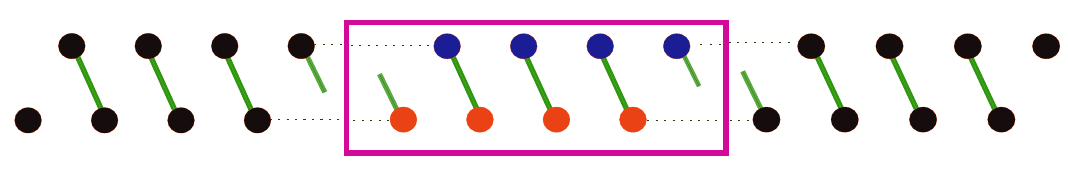}
	\caption{Schematic of the model under an extreme case. This dimerized case of intracellular hopping ($v=0$) reveals that there are two zero energy edge states present in the system.}
	\label{fig:figure2}
\end{figure}
In this case, the edge states are zero energy because we have assumed that the on-site potential at the sandwiching layers is zero.
\subsection{Electronic properties of the system }

To explore information about the nature of edge states, it is illuminating to write the Hamiltonian in matrix form.
\begin{equation}\label{eqn:3}
	\hat{H} = \begin{bmatrix}
		0 & v & 0 & 0 & 0 & 0 & 0  &\dots & 0 \\
		v & 0 & w & 0 & 0 & 0 & 0  &\dots & 0\\
		\vdots& \ddots&\ddots&\ddots&\vdots&\vdots&\vdots&\dots&0\\
		\vdots & \vdots & \ddots & \ddots & \ddots & \vdots & \vdots & \vdots\\
	    0 & \dots & \dots & w & u & v &\dots & \dots & 0\\
	    0 & \dots & \dots & 0 & v & u^{*} & w &\dots &0\\
	    \vdots & \vdots & \vdots & \vdots & \vdots & \ddots & \ddots & \ddots & \vdots\\
	    0 & \dots &\dots & \dots& \dots& \dots & w & 0 &v\\
	    0 & \dots &\dots & \dots& \dots& \dots & \dots & v &0
		\end{bmatrix}.
\end{equation}
The Hamiltonian in Equation 3 is a sparse tri-diagonal matrix. The diagonal element is zero except for the elements from $ (m_{1},m_{1}) $ to $ (m_{2},m_{2}) $, which has stagered potentials $ u $ and $ u^{*} $ representing the non-Hermitian PT symmetric region.

There should be states at the boundaries if the Hamiltonian is associated with a topological property. A look at the spatial spread of the edge-state wave functions confirms their localized nature. The topological nature leads to the fact that the edge states are robust to changes in the Hamiltonian.
\begin{figure}
	\centering
	\includegraphics[width=0.7\linewidth]{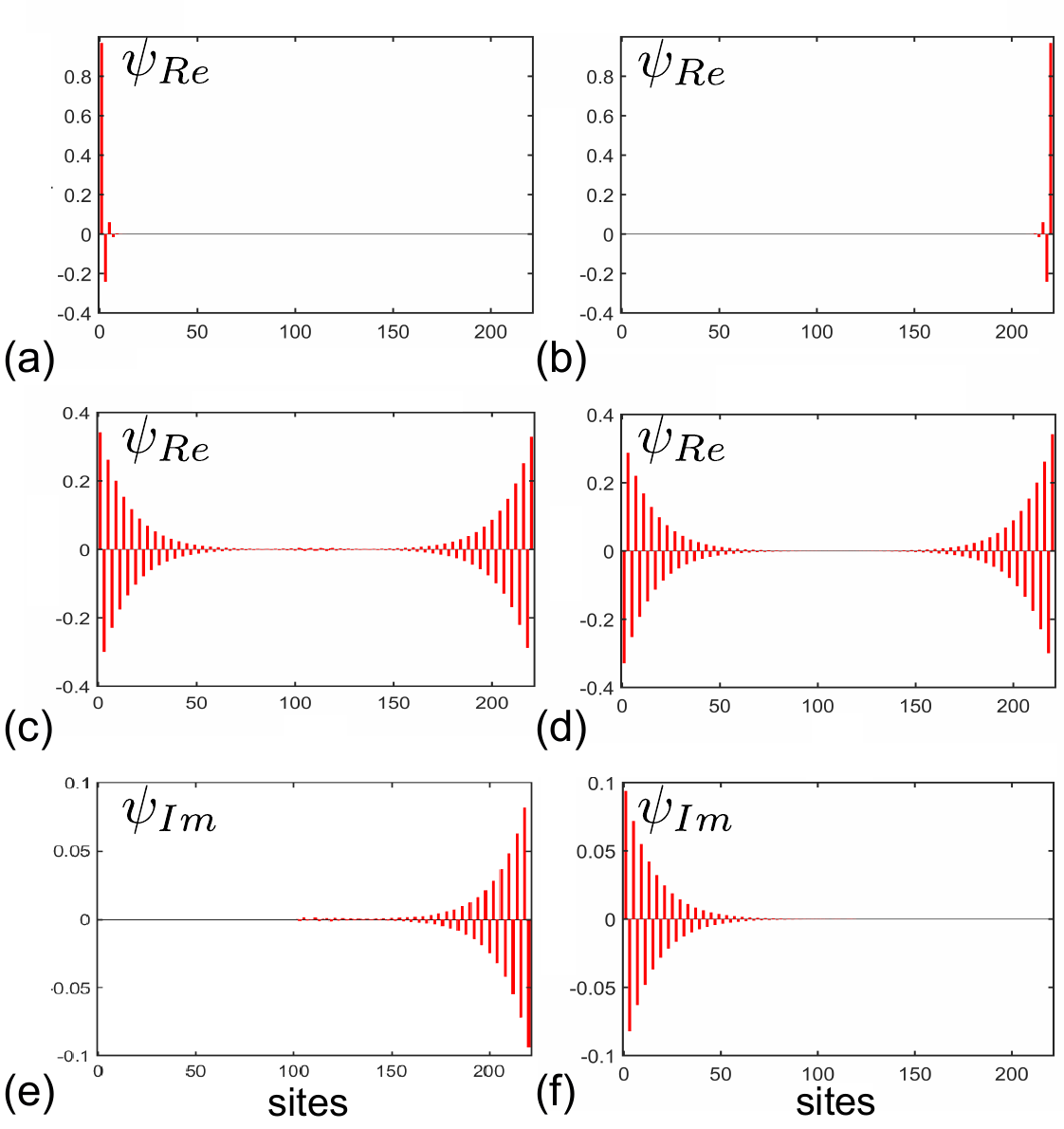}
	\caption{This graph illustrates the spatial spread (amplitude vs. site index) of the two edge-state wavefunctions corresponding to the
		different configurations of the hybrid system of 220 sites with $ u(u^{*}) = -0.3 -(+) 0.1i $ from lattice site $ m_{1}=101 $ to $ m_{2} = 120 $, and $  u = 0 $ elsewhere. (a), and (b) represents
		the real components of the edge states for the case ($v = 0.1$;
		$w = 0:4$), the imaginary components being zero. (c) and (d)
		represents the real part of the wavefunction corresponding to
		($v = 0:35$; $w = 0:4$), whereas (e) and (f) represent the complex part. The tendency of the edge state to delocalise as
		$v \rightarrow 0.4$, with  $w = 0.4$ can be seen in (c--f).}
	\label{fig:figure3}
\end{figure}
It is evident from Figure 3, that the localized nature of the edge-state wavefunctions begin to lose their localized character as we approach the point $v=0.4$, $w=0.4$ from the point $ v = 0.1$, $0.4 $. This reveals the topological nature of the states. Moreover, the overlap of the edge-state wavefunctions of the PT symmetric hybrid system and the plain SSH system, i.e. $ \bra{\psi_{\text{edge}}^{\text{Hybrid}}}\ket{\psi_{\text{edge}}^{\text{SSH}}} $, remains close to $1$ over a range of values of $v$. Hence, the edge state structure of the hybrid system does not change significantly from that of the plain SSH system.
In order to learn more about the nature of the edge states, we need to examine the band structure of the system.
\begin{figure}
	\centering
	\includegraphics[width=0.7\linewidth]{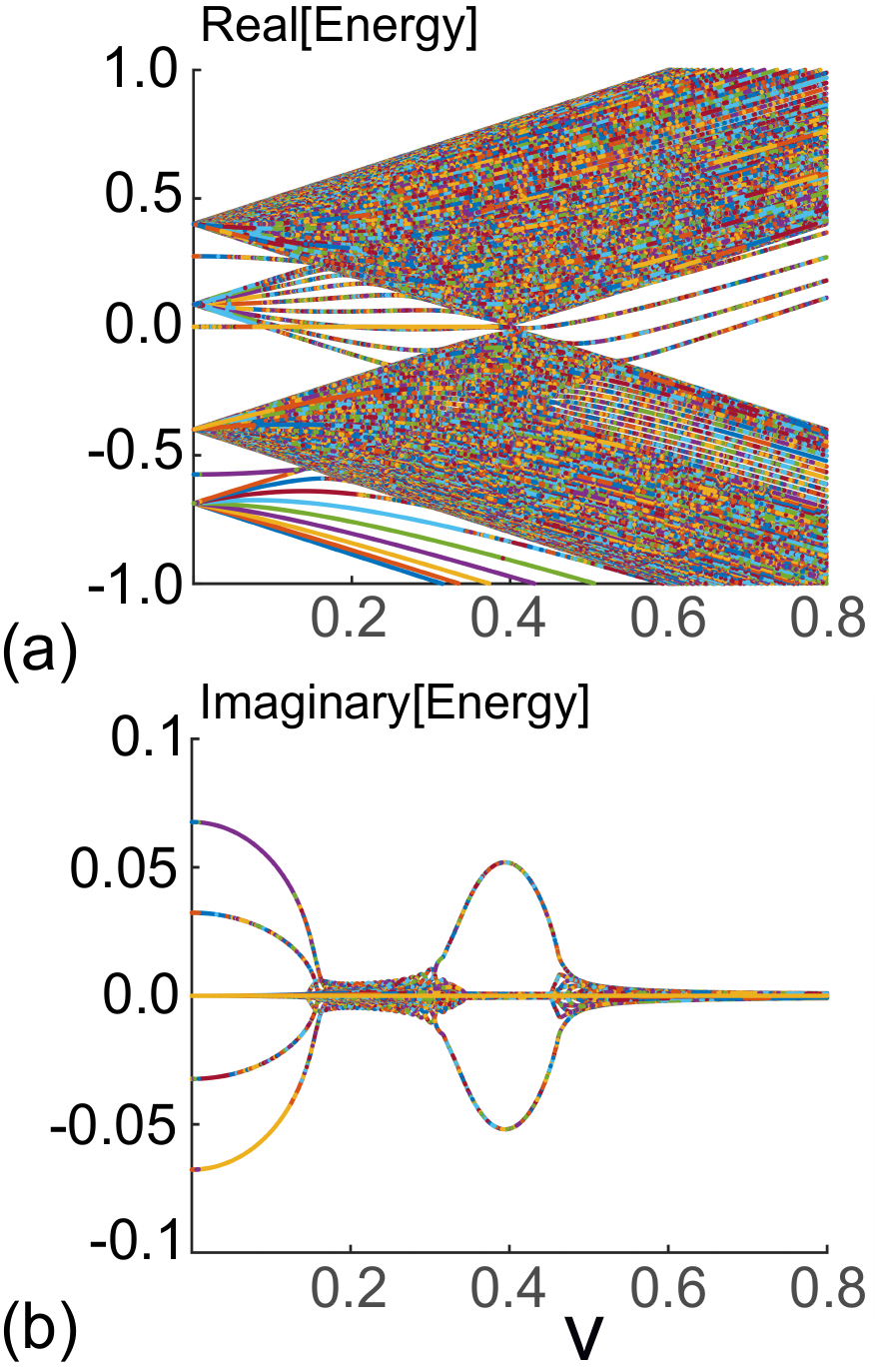}
	\caption{ shows the Energy vs. the intra-cellular hopping parameter $v$. (a) and (b) shows the real and imaginary component of the energy, respectively. The inter-cellular hopping is fixed at $ w=0.4 $. Among lattice sites in the $ m_{1}=101 $ to $ m_{2}=120 $ range, $ u(u^{*})= -0.3-(+0.1i) $, and u=0 for cells outside this range. There is the presence of clearly defined zero energy edge states on the real energy axis till the mark $ v= 0.4 $, on crossing which, the states deviate from their zero energy nature. }
	\label{fig:4}
\end{figure}
In spite of having somewhat unusual characteristics due to the hybrid PT symmetric nature, the energy diagrams do reveal the presence of the robust zero energy edge states, characterizing the topological nature. The real component of the energy shows a clear signature of topological transition at $ v=0.4 $, i.e, for the real part of the energy, there is a band closure at the point when the zero energy states losses its character. The imaginary part of the energy at this point raises from its near zero value before again going to near zero value. This resembles the energy characteristics of a non-Hermitian PT symmetric SSH system in the context of exceptional point~\cite{lieu2018topological}. 
We now proceed forward to study the quench dynamics in this system.
	\section{Results and Discussions \label{sec:3}}
	\subsection{Initial state localization dependent post quench transport}

We have seen in Section 2 that the zero energy states are localized in the ends of the chain i.e, the right end and/or the left end. Now, a quench in a SSH system with an edge state as the initial state $ \psi_{i} $ from a regime (characterized by parameter say $ \chi_{1} $) that holds edge states to a regime (characterized by parameter say $ \chi_{2} $) that is without edge states leads to a post quench transport of the edge state \cite{PhysRevE.108.034102}.  
\begin{equation}\label{eqn:4}
\ket{\psi_{i}(t)}=\sum_{f=1}^{N}e^{-iE_{f}(\chi_{2})t}\ket{\psi_{f}(\chi_{2})}\bra{{\psi_{f}(\chi_{2})}}\ket{\psi_{i}(\chi_{1})}.
\end{equation} 
This transport which essentially is a result of quantum evolution (Equation 4) of states, is from the side which holds the initial edge state to the other side. The system is modeled in such a way that the post-quench probability density wave will face the PT-symmetric region in the middle in a direction that is opposite for the two cases of initial edge states. The middle PT-symmetric layer suggests some asymmetry in the transport characteristics for the two kinds of transport.
\begin{figure}
	\centering
	\includegraphics[width=1.0\linewidth]{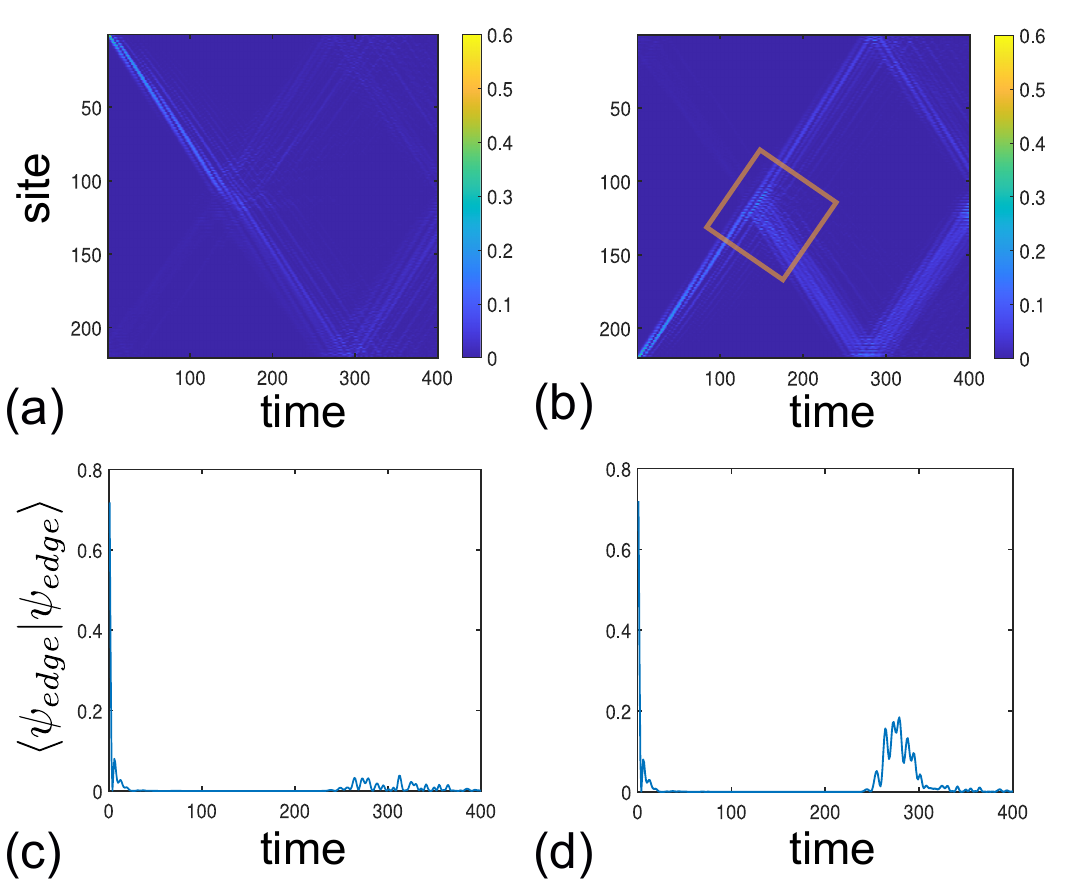}
	\caption{Demonstration of the asymmetry in quench dynamics based on the initial state localization. The hybrid system with the PT-symmetric non-Hermitian region having the alternating potentials $u(u*) = -0.3-(+)0.1i$  from $n1= 101$ to $ n2=120 $ is considered. The quench is performed by the sudden change of the intra-cellular hopping $ v $, from $ 0.1 $ to $ 0.5 $, keeping the inter-cellular hopping $ w $ fixed at $ 0.4 $.  (a) and (b) are the light cone diagram corresponding to the post quench transport, taking the initial edge state to be localized at the left and right end respectively. The marked branching in (b), around the non-Hermitian region from $ m_{1}=101 $ to $ m_{2}=120 $, demonstrates a higher reflection as compared to (a). (c) and (d) are the plots of probability density with respect to the time at the left and the right end, respectively. The magnitude of the re-emergence of the probability density after the initial drop, which indicates the amount of reflection present, is significantly higher in (d) than in (c). }
	\label{fig:5}
\end{figure}
It is interesting to note that Figure. 5 reveals a significant aspect of post-quench transport. The post-quench transport of the probability density displays a stark asymmetry in reflection from the middle PT symmetric layer depending upon the direction in which it is traveling. If the initial edge state is localized on the right-hand side of the system(Figure. 3(b)), i.e., the post-quench transport is towards the left side, then there is a significant reflection of the probability density from the middle layer(see Figure 5(b)), on the contrary, if the initial edge state is the one as shown in Figure. 3(a) that leads to transport of the probability density from the left edge to the right edge, the reflection from the middle layer is almost zero(as shown in Figure 5(a)), similar to unidirectional invisibility as observed in \cite{lin2011unidirectional}. The light cone diagrams in Figures 5(a) and 5(b) correspond to the two cases of post-quench transport in which a sub-cone due to the reflection is prominent in Figure 5(b), and it is nearly absent in 5(a), denoting a reflection-less transport. 
Measuring the probability density with time at the end site where the initial state was initially localized elegantly indicates the magnitude of reflection as is depicted in Figure 5(c) and 5(d). The region of the dip in the probability density corresponds to the time when the probability density is traveling, and the re-emergence of it after a particular time denotes the scenario when the probability density, which is reflected back from the middle layer, has reached the end. By the magnitude of re-emergent probability densities, we are able to quantify the reflection amplitude, and as a result, it is evident that the reflection amplitude is negligible for 5(c) and is substantial for 5(d).
We will now proceed to explain this asymmetrical quench dynamics using transfer matrix method \cite{MarkosSoukoulis+2008, walker1992transfer}.
\subsection{A transfer matrix analysis.}
The post quench transport is reflection-less if the system is entirely an SSH system, i.e., without a middle PT-symmetric non-Hermitian region. It is the introduction of the middle PT-symmetric non-Hermitian layer which gives rise to the scattering of the traveling probability density. Modeling the system for transfer matrix calculation requires consideration of the alternating on-site potentials $ u $ and $ u^{*} $ in the middle layer and the hopping amplitudes associated with them.  Figure 6(a) shows the schematics of modeling the unit potential block $AB$ for transfer matrix analysis;  there is a uniform real potential $ u_{Re} $ spread over the two sites, and the complex potential  $i u_{Im} $ has opposite polarities and different widths $ L_{A} $ and $ L_{B} $ for A and B sites, respectively.
The whole scattering part model, as shown in Figure 6(b), is thus a periodic repetition of the individual potential blocks in Figure 6(a) The transfer matrix is given by, 
\begin{equation}
	T =  [t_{R}]^{-1}[\tilde{T}_{A}][\tilde{T}_{B}]....[\tilde{T}_{A}][\tilde{T}_{B}][t_{L}]
\end{equation}
where, 
\begin{equation}\label{eqn:10}
	\tilde{T}_{\text{A(B)}} = \begin{bmatrix}
		1 & 1\\
		k^{\text{A(B)}}_{+} & k^{\text{A(B)}}_{-}
	\end{bmatrix}\begin{bmatrix}
		e^{ik_{+} L_{\text{A(B)}}} & 0\\
		0 & e^{ik_{-}L_{\text{A(B)}}}
	\end{bmatrix} \begin{bmatrix}
		1 & 1\\
		k^{\text{A(B)}}_{+} & k^{\text{A(B)}}_{-}\end{bmatrix}^{-1}
\end{equation}
The matrix $ \tilde{T}_{\text{A(B)}} $ constitutes information corresponding to the potential block $ \text{A(B)} $. 
It is important to note that, here, $ t_{R} $ and $ t_{L} $ correspond to the potential free regions at right and left ends of the stack, respectively. If T has the form  $ T = \begin{bmatrix}
	\alpha & \beta\\
	\gamma & \sigma
\end{bmatrix} $, then the sacttering matrix will be,
\begin{equation}\label{eqn:7}
	S = \begin{bmatrix}
		-\sigma^{-1}\gamma & \sigma^{-1}\\
		\alpha-\beta\sigma^{-1}\gamma & \beta\sigma^{-1}
	\end{bmatrix}
\end{equation}

\begin{figure}
	\centering
	\includegraphics[width=0.7\linewidth]{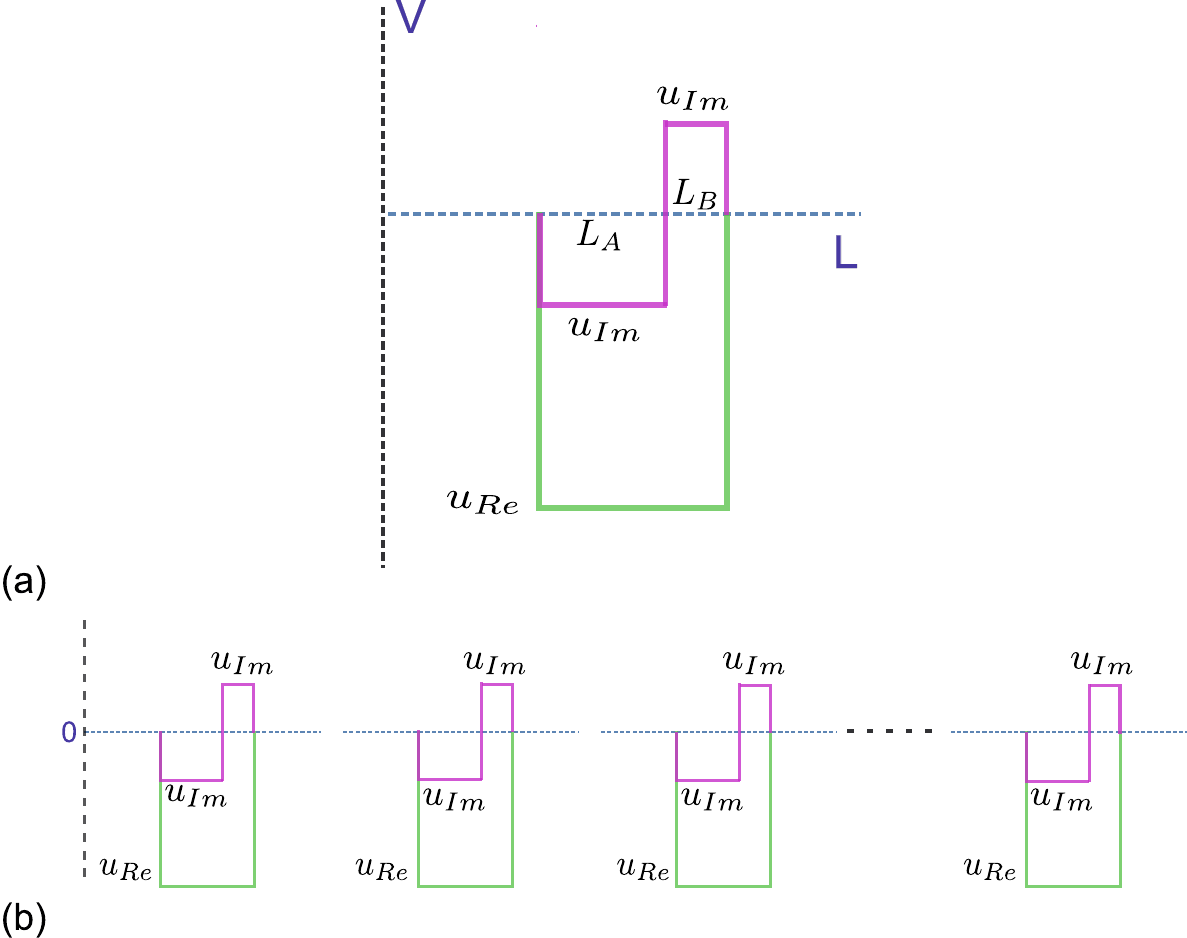}
	\caption{This figure illustrates the schematics of the potential block modelling for the middle PT symmetric non-Hermitian part employed in the transfer matrix calculations. (a) shows the spatial variation of the potential components of each unit block $ AB $. There are two imaginary potential components with reversing polarity,$ -u_{Im} $ and $ u_{Im} $, and spatial spread $ L_{A} $ and $ L_{B} $, respectively, corresponding to the two sub-lattices A and B for the middle scattering part. There is also a uniform real potential($ u_{Re} $) spread through out the two sub-lattices A and B (b) The stacking structure of the potentials represents the whole PT-symmetric middle layer. This diagram is not to scale.}
	\label{fig:6}
\end{figure}

 In Equation 7 for $ S $, the diagonal elements $ S_{11} $ and $ S_{22} $ represents the reflection coefficients for the transport probability wave coming from left and right respectively. 
\begin{figure}
	\centering
	\includegraphics[width=0.7\linewidth]{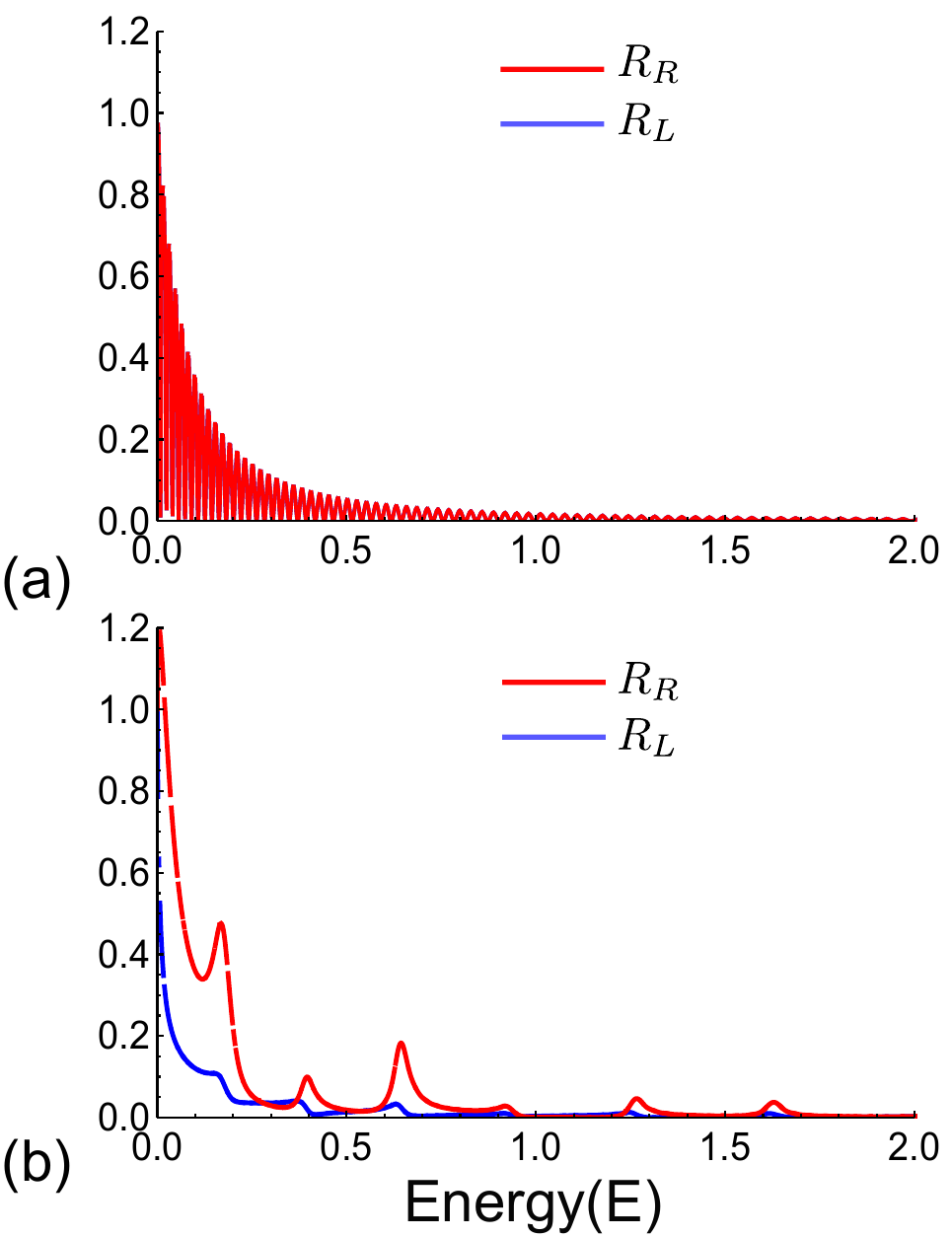}
	\caption{Demonstration of the left and right reflection amplitudes, $ R_{L} $ and $ R_{R} $, arising out of transfer matrix calculations corresponding to $ 10 $ units of potential block $AB$ for the cases when, (a) the scattering part is hermitian with uniform on-site potentials $ u = -0.3 $ and, (b) the scattering part is PT-symmetric non-hermitian with the alternating on-site potential $ u(u^{*})=-0.3-(+)0.1i$ and $ L_{A(B)}=6(10) $. In (a), the reflection amplitudes $ R_{L} $ and $ R_{R} $ are the same, and hence, only one color is prominent as one overlaps the other.}
	\label{fig:7}
\end{figure}
By considering the periodic repetition of potential block $ AB $ as a model for the scattering medium, Figure 7(b) shows that the transfer matrix analysis leads to asymmetric left and right reflection amplitudes ($ R_{L} =|S_{11}|^2$ and $ R_{R}=|S_{22}|^2$) of the transport probability density. Figure 7(a), as a reference, shows that a purely real scattering part without the alternating imaginary component($ u_{Im}=0 $) leads to perfectly symmetric left and right ($ R_{L} $ and $ R_{R} $) reflections, as one might have expected. It is important to note that, in our transfer matrix calculation, this unidirectionality phenomenon arises even under the approximation that the incoming wave is a simple plane wave with particular energy. In the actual post-quench transport considered, this probability wave consists of a series of energy values depending upon the quench, as can be understood from Equation 4.
\section{Summary and conclusions}
We considered a hybrid model designed to exhibit highly anisotropic post-quench transport. The system comprises a non-Hermitian PT-symmetric SSH system sandwiched between two ordinary SSH systems. This system is the first demonstrated to have topological characteristics and is able to host robust, well-defined edge states. An extreme dimerized case was used to demonstrate the presence of edge states geometrically, and then the spatial structure of the edge states was studied, and it was found that the localization character of the edge state is maintained for a range of hopping parameters, revealing its topological character. The interesting character of the states is revealed by examining the system's energy vs. the parameters, which included zero energy states and exceptional point like behaviour. 

We then proceeded forward to study the post-quench transport. The study of the "light cone" diagrams reveals the directional dependence of the post-quench transport. The direction of the transport of the probability density depends upon the end at which the initial wave -function was localized. It is found that in one direction, there is almost no reflection of the traveling probability density wave, whereas, in the other direction, there is a considerable reflection from the middle PT-symmetric region. In order to a deeper understanding of this unidirectional post-quench transport, we model the system with an assembly of potentials ordered in a repeated fashion to mimic the middle PT-symmetric part of our hybrid system. Using transfer matrix calculations, it is apparent that there is an asymmetry between left and right reflection amplitudes. In cases where the energy is too large or too low, both of the reflection amplitudes approach the same value. Summarily, the post-quench transport for this hybrid system depends on the initial edge state localization, i.e., the initial localization (right end or left end) determines the direction of post-quench transport, which determines quench dynamics.
\section*{ACKNOWLEDGMENTS}
     Melbourne Research Scholarship and N.D Goldsworthy Scholarship fund this project. We thank the University of Melbourne and the Indian Institute of Technology Kharagpur for providing a supportive research environment.

\bibliographystyle{unsrtnat}
\bibliography{References}

\end{document}